\documentclass[prd,aps,letterpaper,twocolumn,superscriptaddress,preprintnumbers,nofootinbib,floatfix]{revtex4}
\pdfoutput=1
\usepackage{color}
\usepackage[pdftex]{graphicx}
\usepackage{amsmath}
\usepackage{amsfonts}
\usepackage{amssymb}
\usepackage{rotating}
\usepackage{subfigure}
\usepackage{paralist} 
\usepackage{verbatim}
\usepackage{float}
\usepackage{soul} 
\usepackage[title]{appendix}
\usepackage{natbib}
\usepackage{epsfig}

\usepackage[utf8]{inputenc}
\usepackage[english]{babel}
\usepackage{tabularx}
\newcolumntype{C}{>{\centering\arraybackslash}X}
\usepackage{mathpazo} 
\usepackage{tgpagella} 
\usepackage[colorlinks=true,linkcolor=red,urlcolor=blue,citecolor=blue]{hyperref}
\newcommand{\stonybrook}{Physics and Astronomy Department, Stony Brook University, Stony Brook, NY  11794}

\newcommand{\fgsnrmissub}{$1.25 \sigma$}
\newcommand{\fgsnrcibmissub}{$1.11 \sigma$}
\newcommand{\fgsnrtszmissub}{$0.59 \sigma$}
\newcommand{\fdmsnrmissub}{$5.06 \sigma$}
\newcommand{\fdmsnrmissubdiagonly}{$9.37 \sigma$}
\newcommand{\fdmsnranalyticdiag}{$13.42 \sigma$}
\newcommand{\fdmsnr}{$5.11 \sigma$}
\newcommand{\fgsnr}{$1.73 \sigma$}

\begin{document}

\title{Mitigating Foreground Bias to the CMB Lensing Power Spectrum for a CMB-HD Survey}
\author{Dongwon Han}
\affiliation{\stonybrook}

\author{Neelima~Sehgal}
\affiliation{\stonybrook}

\begin{abstract}
A promising way to measure the distribution of matter on small scales ($k \sim 10~h$Mpc$^{-1}$) is to use gravitational lensing of the Cosmic Microwave Background (CMB).  CMB-HD, a proposed high-resolution, low-noise millimeter survey over half the sky, can measure the CMB lensing auto spectrum on such small scales enabling measurements that can distinguish between a cold dark matter (CDM) model and alternative models designed to solve problems with CDM on small scales.  However, extragalactic foregrounds can bias the CMB lensing auto spectrum if left untreated. We present a foreground mitigation strategy that provides a path to reduce the bias from two of the most dominant foregrounds, the thermal Sunyaev-Zel'dovich effect (tSZ) and the Cosmic Infrared Background (CIB). Given the level of realism included in our analysis, we find that the tSZ alone and the CIB alone bias the lensing auto spectrum by $0.6\sigma$ and $1.1\sigma$, respectively, in the lensing multipole range of $L \in [5000,20000]$ for a CMB-HD survey; combined these foregrounds yield a bias of only $1.3\sigma$.  Including these foregrounds, we also find that a CMB-HD survey can distinguish between a CDM model and a $10^{-22}$~eV FDM model at the $5\sigma$ level.  These results provide an important step in demonstrating that foreground contamination can be sufficiently reduced to enable a robust measurement of the small-scale matter power spectrum with CMB-HD.
\end{abstract}

\maketitle

\section{Introduction}
\label{sec:intro}

The nature of dark matter (DM) is a major open question in cosmology and particle physics. In the context of cosmology, a cold dark matter (CDM) model, where dark matter consists of a non-relativistic, collisionless particle, yields predictions that are consistent with observations of the large scale structure (e.g.~\cite{Peebles1982, Blumenthal1984, Davis1985, Frenk2012, Primack2012, DES2021}). However, on smaller scales less than 10 kpc (corresponding to $M<10^9 M_{\odot}$), CDM may overestimate the amount of structure (e.g~\cite{Flores1994, Moore1994, Klypin1999, Moore1999, Ostriker2003,BoylanKolchin2011,Brooks2014, Weinberg2015, Oman2015, DelPopolo2017, Bullock2017}). Many alternative models of dark matter have been proposed to solve this issue. These alternative models include warm dark matter (WDM)~\cite{Colin2000,Bode2001,Viel2005}, fuzzy dark matter (FDM)~\cite{Turner1983, Press1990, Sin1994,  Goodman2000, Hu2000a, Peebles2000,Amendola2006, Schive2014, Marsh2016b, Hui2017}, self-interacting dark matter (SIDM)~\cite{Carlson1992, Spergel2000, Vogelsberger2012, Fry2015, Elbert2015, Kaplinghat2016, Kamada2017, Huo2018, Tulin2018}, and superfluid dark matter (SFDM)~\cite{Berezhiani2015,Khoury2021}, to name a few. These alternative DM models predict varying amounts of the suppression of structure growth on small scales. In addition, baryonic processes, such as AGN feedback, can alter the distribution of matter on small scales~(e.g.~\cite{VanDaalen2011, Brooks2013, Brooks2014, Natarajan2014}). Recent work suggests that baryons alter the shape of the matter power spectrum in a finite set of ways that can be described with a small set of parameters~\cite{Schneider2019,Giri2021}; moreover, alternate dark matter models appear to change the shape of the matter power spectrum in a way that differs from baryonic processes~\cite{Nguyen2019}.  Therefore, a robust measurement of the matter power spectrum below 10~kpc scales will impose stringent constraints on alternative dark matter models and baryonic effects, and has the potential to distinguish between the two. 

One promising avenue to measure small-scale structure is a high-resolution (0.25~arcminute), low-noise ($0.5~\mu K$-arcmin) CMB lensing measurement~\cite{Nguyen2019, Sehgal2019a, Sehgal2019b,Sehgal2020}. Unlike baryonic tracers, such as a galaxy number counts or Lyman-$\alpha$, CMB lensing measurements probe the underlying dark matter distribution directly through gravitational lensing. Furthermore, in a cold dark matter paradigm, lower mass halos formed first at high redshift. Since the CMB lensing kernel peaks at relatively high redshift ($z \approx 2$), it is, in principle, more sensitive to the suppression of structure than local probes.  CMB-HD, an ultra-deep, high-resolution millimeter-wave survey over half the sky, has the potential to measure the CMB lensing auto spectrum on small scales of order $k \sim 10~h$Mpc$^{-1}$, which no precursor CMB survey can achieve~\cite{Sehgal2019a, Sehgal2019b, Sehgal2020}. In particular, CMB-HD is sensitive enough to make a robust measurement of the small-scale matter power spectrum, and determine whether baryonic physics or modifications to CDM, or potentially both, are needed to match observations of the matter distribution on small scales.

The CMB lensing signal on small scales gains most of its signal-to-noise ratio (SNR) from CMB temperature maps~\cite{Nguyen2019}. However, CMB temperature maps include contamination from extragalactic foregrounds, such as the thermal and kinetic Sunyaev-Zel’dovich effects (tSZ and kSZ), the Cosmic Infrared Background (CIB), and radio galaxies (Radio). If untreated, these extragalactic foreground components can significantly bias the CMB lensing signal~\cite{VanEngelen2014, Osborne2014, Ferraro2018, Schaan2019, Sailer2020, Sailer2021}. Many techniques have been proposed to mitigate this foreground bias, such as gradient-cleaning, foreground deprojection, and biased-hardened lensing estimators~\cite{Madhavacheril2018, Darwish2020, Sailer2020}. However, previous works have mainly focused on the impact of foregrounds on larger scales ($L \le 5000$). In this work, we focus on a mitigation strategy for smaller scales ($L \in [5000, 20000]$), which contribute most of the SNR in distinguishing, for example, alternative DM models from a CDM model~\cite{Nguyen2019}.  We apply these techniques to simulations from~\cite{Sehgal2010} that include realistic non-Gaussian extragalactic foregrounds, after modifying these simulations as described in Section~\ref{sec:sims}.  In Section~\ref{sec:method}, we describe the CMB lensing estimator and foreground mitigation techniques used in this work. In Section~\ref{sec:results}, we show that we are able to reduce the two most dominant foregrounds, the tSZ and CIB, enough to make an unbiased measurement of the CMB lensing auto spectrum in the range of $L \in [5000, 20000]$, and robustly distinguish between interesting dark matter models.  We discuss additional systematic considerations in Section~\ref{sec:discussion}, and conclude in Section~\ref{sec:conclusion}.

\section{Simulations}
\label{sec:sims}

We use the non-Gaussian extragalactic foreground simulations from~\cite{Sehgal2010} (hereafter S10 simulations), since these simulations have extragalactic foregrounds correlated with each other. We rescale the S10 CIB and tSZ maps by a factor of 0.75 following~\cite{VanEngelen2014, SO2019} to make the S10 simulations more closely match the latest observations~\cite{Dunkley2013, Sievers2013, Planck2014SZ, George2015, Planck2016tSZ}. We cut out four $20^{\circ} \times 20^{\circ} = 400$ square degree patches near the equator, and reproject the corresponding S10 CIB, tSZ, and Radio maps from this patch onto a flat-sky grid with 0.25 arcmin resolution to match the resolution of CMB-HD at 150 GHz. We call these maps the S10 CIB, tSZ, and Radio patch maps. 

\subsection{Creation of High-Resolution Lensing Potential Map Correlated with tSZ and CIB}

The correlation between the CIB and the lensing convergence map, $\kappa$, in the S10 simulations is roughly $35\%$ at 150~GHz~\cite{Han2021}, which is significantly lower than the 70\% correlation measured subsequently by {\it{Planck}}~\cite{Planck2014CIB}. In addition, the S10 $\kappa$ map, which has a resolution of 1 arcminute, only has the information up to $\ell \approx 10000$, missing the small scale information needed for this analysis. Therefore, we replace the S10 $\kappa$ map with one we make ourselves that has higher resolution and $70\%$ correlation with the CIB by construction.  Similarly, the tSZ-$\kappa$ anti-correlation is $45\%$ at 150 GHz in S10, as opposed to roughly $50\%$ found in~\cite{Hill2014tSZ} over the range of $\ell \in [100,10000]$.  Thus we increase the correlation between the S10 tSZ map and our new $\kappa$ map to $50\%$. In this work, we focus on the bias from the tSZ and CIB to the lensing signal, as these foregrounds generally have the largest contributions to the lensing bias~\cite{VanEngelen2014,Han2021}; thus we do not explicitly include kSZ-$\kappa$ and Radio-$\kappa$ correlations in the construction of the new $\kappa$ map. However, we do include the kSZ and Radio signals as additional sources of noise throughout this work.  We leave an indepth exploration of kSZ and Radio bias to the lensing signal for future work.   

We model this new $\kappa$ map as a sum of two parts; one that is correlated with the foregrounds and one that is independent of the foregrounds. We create this new $\kappa$ map following Equation~\ref{eq:kappasim} below.

\begin{widetext}
\begin{align}\label{eq:kappasim}
\begin{split}
\kappa &=  M_{G} + \frac{1}{1-[\rho^{CIB \times tSZ}(\ell)]^2}[\rho^{\kappa \times CIB}(\ell)-\rho^{CIB \times tSZ}(\ell)\rho^{\kappa \times tSZ}(\ell)]\sqrt{\frac{C_{\ell}^{\kappa \kappa}}{C_{\ell}^{CIB }}}M_{CIB}^{S10} \\ 
&+  \frac{1}{1-[\rho^{CIB \times tSZ}(\ell)]^2}[\rho^{\kappa \times tSZ}(\ell)-\rho^{CIB \times tSZ}(\ell)\rho^{\kappa \times CIB}(\ell)]\sqrt{\frac{C_{\ell}^{\kappa \kappa}}{C_{\ell}^{tSZ }}}M_{tSZ}^{S10} \\
\end{split}
\end{align}
\end{widetext}
Here, $C_{\ell}^{\kappa \kappa}$ is a theory lensing convergence power spectrum generated using CAMB~\cite{Lewis2000}. $M_{CIB}^{S10}$ and $M_{tSZ}^{S10}$ are the S10 CIB and tSZ patch maps discussed above. We calculate $C_{\ell}^{CIB}$ and $C_{\ell}^{tSZ}$ by measuring the power spectrum of the full-sky S10 CIB and tSZ maps at 150~GHz. These raw spectra are noisy due to the single realization of the S10 simulation set; thus, we smooth the resulting power spectra with a $20\sigma$ Gaussian kernel, which smooths the spectra over $\Delta \ell \sim 20$, to reduce this noise. Similarly, we compute the correlation coefficient, $\rho^{CIB \times tSZ}(\ell)=C_\ell^{CIB \times tSZ}/\sqrt{C_\ell^{CIB} C_\ell^{tSZ}}$, using the full-sky S10 simulations. We set $\rho^{\kappa \times CIB}(\ell)$ and $\rho^{\kappa \times tSZ}(\ell)$ 
to 0.7 and -0.5 following the discussion above. Finally, we generate a Gaussian random field, $M_G$, such that $C_{\ell}^G = \langle  M_G, M_G \rangle = C_{\ell}^{\kappa \kappa} - C_{\ell}^{\kappa \times FG}$, where we define $C_{\ell}^{\kappa \times FG}$ as the cross spectra between $\kappa$ and the combined tSZ plus CIB foregrounds; $M_G$ serves as the part of the $\kappa$ map uncorrelated with the foregrounds. In the limiting case where we have only one correlated foreground (e.g.,~let  $M_{tSZ}^{S10}=0$), Equation~\ref{eq:kappasim} simplifies to Equation 3.2 from~\cite{Kamionkowski1997}. In this simplified case, it is apparent that Equation~\ref{eq:kappasim} is constructed to give the correct cross power spectrum between $\kappa$ and the correlated foreground. 

Since we are constructing $\kappa$ as a scaled combination of $M_{CIB}^{S10}$ and $M_{tsz}^{S10}$, we include the maximum amount of non-Gaussian features from these foregrounds in our $\kappa$ maps. Therefore, higher-order correlations between these new $\kappa$ maps and the foreground components are expected to be larger than those found in the original S10 $\kappa$ simulations. In addition, a higher correlation between $\kappa$ and foregrounds typically leads to a more significant foreground bias.  In these respects, we expect the bias we find with these new $\kappa$ maps to be conservative.

Since there is a small difference between the power spectra of the CIB and tSZ patch maps ($M^{S10}_{CIB}$ and $M^{S10}_{tSZ}$) and the full-sky average spectra used in Equation~\ref{eq:kappasim}, we find that the new $\kappa$ maps have a slight excess of power compared to the expected theory power spectrum, $C_{\ell}^{\kappa \kappa}$. Thus, for each of the four 400 square degree patches, we compute a transfer function by taking the ratio of the average power spectrum of 30 independent $\kappa$ simulations and the CAMB theory $C_{\ell}^{\kappa \kappa}$. We find that the transfer function is less than 1 percent of the theory power between $\ell=2000$ and $\ell=28000$. We apply this transfer function to the final version of $\kappa$ maps used in this analysis. In Appendix~\ref{sec:gvsngkappa}, we show an image of one of our non-Gaussian $\kappa$ maps, and compare it to a Gaussian $\kappa$ map generated with the same $C_{\ell}^{\kappa \kappa}$. We generate 25 realizations of these $\kappa$ maps for each of the 400 square degree patches. With these new $\kappa$ maps in hand, we generate lensed CMB maps by first making 100 Gaussian unlensed CMB realizations of the same footprint as the foregrounds, using a theory spectra generated with {\it{Planck2018}} cosmological parameters~\cite{Planck2018Parameters}. Then, we lens the unlensed CMB maps with the new $\kappa$ maps using the publicly available {\it{pixell}} library\footnote{\href{https://github.com/simonsobs/pixell}{https://github.com/simonsobs/pixell}\label{fn:pixell}}. 

\begin{figure*}[t]
  \centering
  \includegraphics[width=0.49\textwidth]{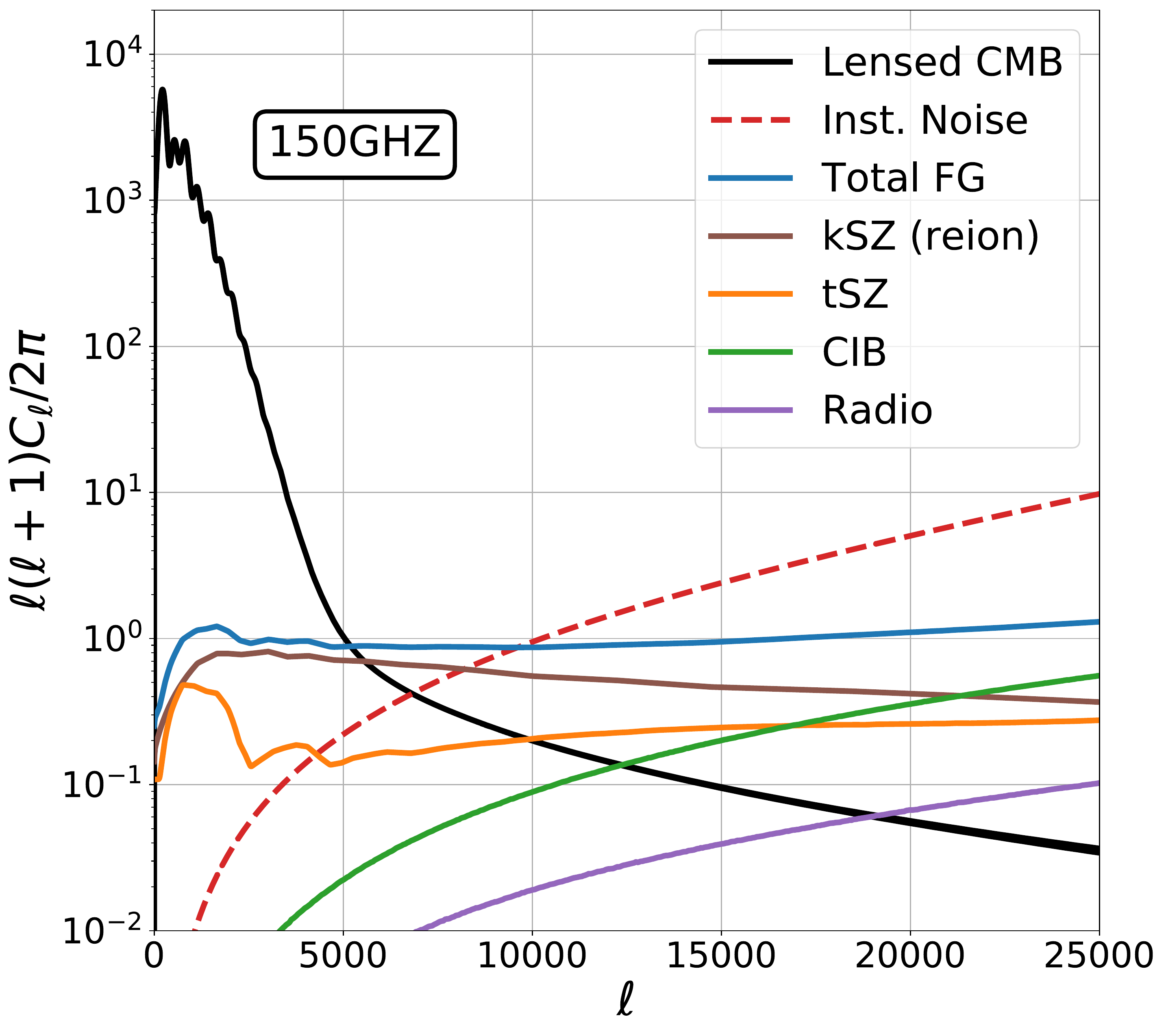}
  \includegraphics[width=0.49\textwidth]{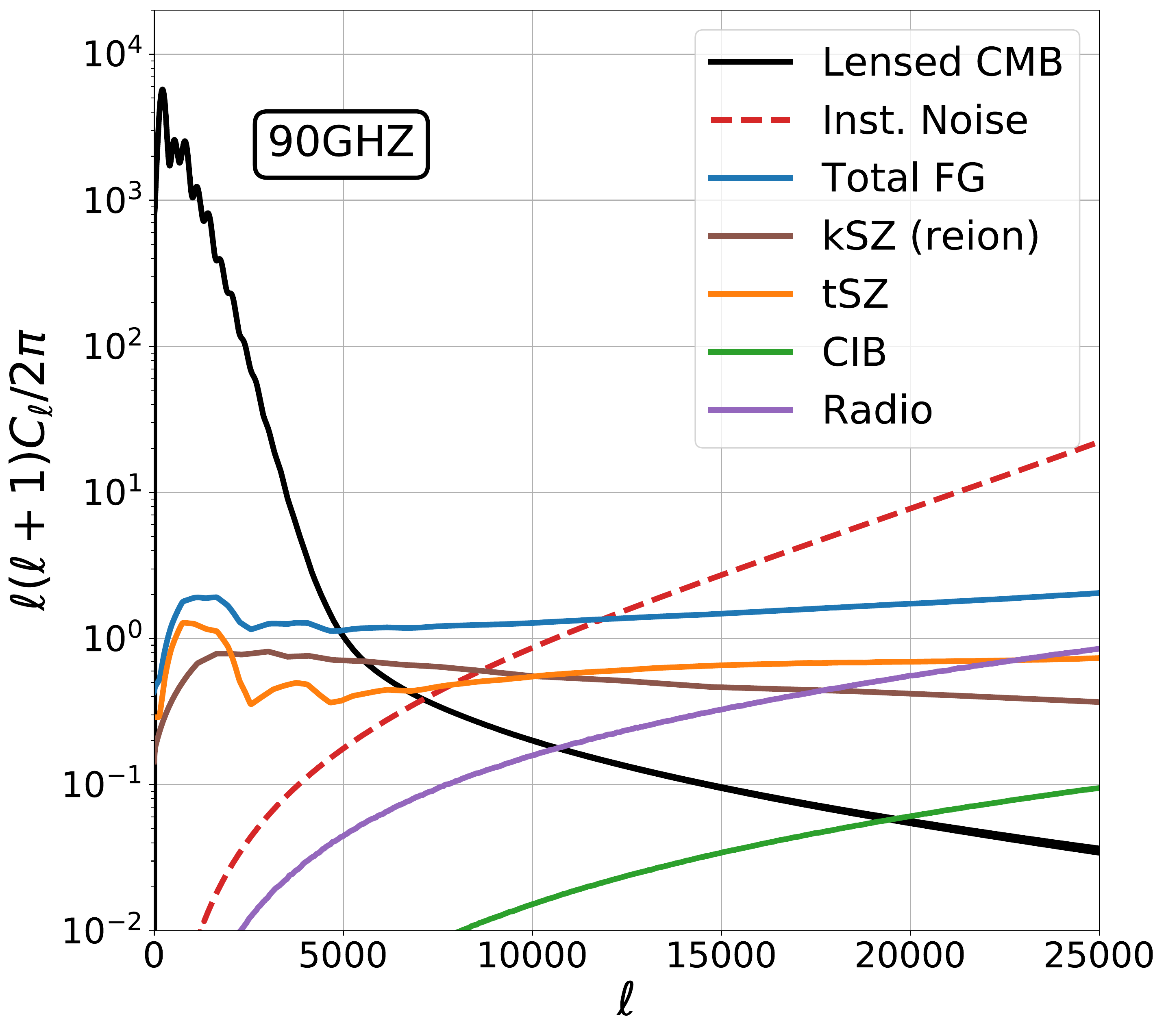}
  \caption{Shown are the expected power spectra at 150~GHz ({\it{left}}) and 90 GHz ({\it{right}}) for a CMB-HD survey. The theoretical CMB power spectra are calculated using CAMB~\cite{Lewis2000} with {\it{Planck2018}} cosmological parameters~\cite{Planck2018Parameters}. Instrumental noise curves are shown as red dashed curves. CMB-HD is expected to have a resolution of 0.25 and 0.42 arcminutes, and a sensitivity of 0.8 and 0.7 $\mu$K-arcmin noise, in the 150 and 90~GHz channels, respectively~\cite{Sehgal2020}. The reionization kSZ spectrum is shown as the brown curve~\cite{Smith2018}. The tSZ, CIB, and Radio power spectra are computed using the S10 simulations, with the scalings and source cuts as described in Section~\ref{sec:sims}. In particular, we assume CIB sources above 0.03 mJy and 0.008 mJy at 150 GHz and 90 GHz, respectively, are removed, as well as Radio sources above 0.03 mJy and 0.04 mJy at at 150 GHz and 90 GHz, respectively. Note that we also account for some amount of CIB mis-subtraction due to uncertainty in the CIB spectral index for source fluxes extrapolated from higher frequencies, as discussed in Section~\ref{sec:sims}. We also assume that tSZ clusters detected at the $3\sigma$ level are removed from the maps. The total foreground spectra (blue curves) are the sum of the reionization kSZ, tSZ, CIB and Radio spectra.}\label{fig:spec}
\end{figure*}

\subsection{Creation of Foreground-Reduced Maps}

We also make foreground-reduced versions of the CIB, tSZ, and Radio patch maps, with flux and cluster cuts matching what we expect to be achieved by CMB-HD.  CMB-HD is expected to have a resolution of 0.25 and 0.42 arcminutes, and a sensitivity of 0.8 and 0.7 $\mu$K-arcmin white noise, for the 150 and 90~GHz frequency channels, respectively~\cite{Sehgal2019b,Sehgal2020}. Given these noise levels and resolution, we expect to detect point sources (i.e.~galaxies) above 0.04 mJy at 90 GHz with more than $5\sigma$ significance~\cite{HDWEBSITE}. This detection threshold assumes the absence of source confusion from blended sources, which is a reasonable assumption for Radio sources since they are not so densely clustered; for reference, we expect the number density of radio sources to be less than 0.07 radio source per 0.42 arcminute beam and 0.03 radio source per 0.25 arcmin beam~\cite{Sehgal2010}. The $5\sigma$ detection thresholds given in~\cite{HDWEBSITE} are based on an analytic calculation assuming only white noise, which is the dominant source of noise at the small scales considered here.  Thus we assume that Radio sources can be detected and removed that are above $0.04$ mJy at 90~GHz, and correspondingly $0.03$ mJy at 150 GHz, assuming a spectral index of -0.8.  (We note that there will be some scatter in the extrapolation of radio fluxes from 90 GHz to 150 GHz for source too faint to be detected in the 150~GHz maps directly, and we ignore this scatter in this analysis.)  We assume that the point source removal is done via template subtraction, since the point sources have a known shape corresponding to the instrument beam and their fluxes can be measured; the advantage of template subtraction is that we do not cut holes in the map or disturb the coincident lensing signal. 

For CIB sources, given CMB-HD noise levels and resolution, as well as the confusion of blended CIB sources (which are more numerous and clustered than Radio sources), we find that sources above 0.2 mJy can be detected at the $5\sigma$ level in 150 GHz maps directly~\cite{Prasad2022}.  To get to a lower flux threshold in 90 and 150 GHz maps, we exploit the fact that CMB-HD will have a 280 GHz channel with 2.7 $\mu$K-arcmin white noise.  We find that CIB sources above 0.15 mJy at 280 GHz can be detected at the $5\sigma$ level in the 280 GHz channel; this detection level was determined by applying a matched-filter on maps that included confusion from other CIB sources, as well as the kSZ, tSZ, and CMB~\cite{Prasad2022}.  Assuming a spectral index of 2.6 for CIB sources~\cite{Sehgal2010}, this results in the identification of sources above 0.03 mJy at 150 GHz and 0.008 mJy at 90 GHz. Once detected, we assume these sources can be removed using template subtraction, using a fluxes extrapolated from the 280 GHz map. Since there can be uncertainty on the CIB spectral index for a given source, this can lead to over or under-subtraction of CIB sources at 90~GHz and 150~GHz. Thus in our analysis, we incorporate some amount of mis-subtraction (see below); we find that this mis-subtraction of CIB sources has minimal impact on the lensing bias (see Section~\ref{sec:results} for more details).

To create the foreground-reduced point source maps, with the flux cuts discussed above, we remove sources in the CIB and Radio maps prior to adding them to the lensed CMB maps.
To do this, we find all the pixel locations of CIB sources with fluxes above $0.03$ mJy at 150~GHz, and set them to the mean flux of the remaining CIB map (which includes only sources below 0.03 mJy); this mimics the diffuse CIB that will still reside at the template-subtracted locations.  We follow a similar procedure for the 90 GHz map. To account for some uncertainty in the CIB spectral index for sources whose flux is extrapolated from 280 GHz, we ``mis-subtract" sources with fluxes between 0.03~mJy and 0.2~mJy at 150 GHz by up to 5\% of their original intensity in both the 150 and 90 GHz maps. For the sources in this flux range, we add back to the appropriate locations their original pixel values multiplied by random numbers drawn from $\mathcal{U}[-0.05, 0.05]$.  To remove the Radio sources, we zero out all Radio sources with fluxes above $0.04$ mJy at 90 GHz from both the 90~GHz and 150~GHz maps. We perform these cuts in the S10 patches at 0.5 arcminute resolution, before projecting these maps to 0.25 arcminute resolution; we do this because the S10 model was only matched to observations in its native resolution of 0.5 arcminute, and may not necessarily match observations if we extrapolate it to higher resolutions than it was intended to model.{\footnote{The S10 simulations include a catalog of CIB sources in which many low flux sources add together to a larger flux source when placed in a 0.5 square arcminute pixel.  If we had instead populated 0.25 arcminute maps with the CIB sources from the catalog, then we would have obtained a higher CIB power spectrum than observations allow.  Thus we make the CIB flux cut in maps at the native S10 resolution of 0.5 arcminute, and then project the residual sources to higher resolution by upsampling the maps.}} Since the Radio sources are not correlated with the lensing potential map by construction, after obtaining the power spectra of the residual Radio sources at each frequency, we model this component as a Gaussian random field.

We also assume the removal of all $3\sigma$ detected tSZ clusters.  These clusters will be found based on their integrated cluster Compton-y parameter, $\textrm{Y}$, which is closely related to cluster mass.  We use the cluster mass threshold for $5\sigma$ detected clusters found for CMB-HD by~\cite{Raghunathan2021} and shown in their Figure 3, and we convert this to a mass threshold for $3\sigma$ detected clusters by multiplying their mass threshold by 0.75, which assumes their Y-M scaling relation of $\textrm{Y} \propto M^{1.79}$.  We assume clusters above this mass threshold are perfectly subtracted from our maps. To simulate the tSZ cluster subtraction, we find the coordinates of the tSZ clusters with masses above the mass threshold using the S10 cluster catalogue. We then mask these tSZ clusters in the tSZ maps by replacing them with 5 and 10 arcminute radius holes for the tSZ clusters with $z > 0.5$ and $z < 0.5$, respectively. Since, a few of the closest tSZ clusters, which have irregular-shapes and large angular diameters ($\approx 1~\textrm{degree}$), are not fully covered by the tSZ masks, we visually find these irregular sources in the tSZ maps and set their pixel values to zeros. Note that the masking is done directly in the tSZ maps, simulating a near-perfect subtraction.  We discuss mis-subtraction in the context of CIB in Section~\ref{sec:results}, and leave an indepth analysis of cluster mis-subtraction to future work.

The kSZ signal consists of two components, 1)~a contribution from the epoch of reionization and 2)~a late-time, lower-redshift, component (hereafter reionization kSZ and late-time kSZ, respectively)~\cite{Smith2018}. We assume that the late-time kSZ can be removed from the large-scale gradient map with an overlapping galaxy survey~\cite{Smith2018,Cayuso2021}, such as from the Rubin Observatory~\cite{LSST2019}, and that it can be removed from the small-scale map with machine learning techniques (e.g.~\cite{Wu2019, Petroff2020, Guzman2021a,Guzman2021b, Han2021, Lin2021, Villanueva2021}) (see Section~\ref{sec:discussion}). The reionization kSZ originates at early times ($z \approx 10$) and is more Gaussian than the lower redshift foregrounds. In addition, the CMB lensing kernel peaks at $z \approx 2$ and thus has limited overlap with the reionization kSZ. Previous work focusing on large scales has found the lensing bias from the reionization kSZ to be negligible for a CMB-S4-type experiment~\cite{Cai2021}; we leave exploration of the lensing bias from the reionization kSZ for a CMB-HD-type experiment to future work.  Here, we approximate the reionization kSZ as a Gaussian random field, which adds noise to the lensing reconstruction but not bias.

In Figure~\ref{fig:spec}, we show the power spectra of our simulations at 150~GHz (left) and 90~GHz (right). The theoretical CMB power spectra are calculated using CAMB~\cite{Lewis2000} with {\it{Planck2018}} cosmological parameters~\cite{Planck2018Parameters}. The instrumental noise spectra are computed following the CMB-HD technical requirements~\cite{Sehgal2020}.  The power spectra of the foregrounds reflect the reduced-foreground model described above.

\section{Method}
\label{sec:method}

A standard way to measure a lensing signal from CMB maps is to reconstruct a lensing convergence map, $\kappa$, using a quadratic estimator ($\mathcal{Q}$)~\cite{Hu2000b, Hu2002, Sherwin2017, Planck2018Lensing, SPT2019}. In the presence of non-Gaussian foregrounds correlated with a $\kappa$ map, the reconstructed lensing auto spectrum can be decomposed into four terms: 1)~the lensing signal, 2)~the primary bispectrum bias, 2)~the secondary bispectrum bias, and 4)~the trispectrum bias~\cite{VanEngelen2014, Osborne2014, Ferraro2018, Schaan2019, Sailer2020, Sailer2021}. Modeling the signal ($S$) as the sum of the lensed CMB temperature map ($T$) and the extragalactic foregrounds ($FG$) (i.e., $S = T + FG$), we have

\begin{align}\label{eq:lens_auto}
\begin{split}
\underbrace{\langle \mathcal{Q}[S,S], \mathcal{Q}[S,S] \rangle}_\text{lensing auto spectrum}  &= \underbrace{\langle \mathcal{Q}[T,T], \mathcal{Q}[T,T] \rangle}_\text{lensing signal} \\
&+\underbrace{2\langle \mathcal{Q}[T,T], \mathcal{Q}[FG,FG] \rangle}_\text{primary bispectrum bias} \\
&+\underbrace{4\langle \mathcal{Q}[T,FG], \mathcal{Q}[T,FG] \rangle}_\text{secondary bispectrum bias} \\
&+\underbrace{\langle \mathcal{Q}[FG,FG], \mathcal{Q}[FG,FG] \rangle}_\text{trispectrum bias}
\end{split}
\end{align}

In this work, we use the quadratic estimator described in~\cite{Hu2007} (hereafter HDV estimator), which is tailored to reconstruct $\kappa$ maps on small scales. On small scales, lensing induces small dipole perturbations that are aligned with the smoothly varying background primordial CMB gradient~\cite{Dodelson2003, Holder2004, Lewis2006, Madhavacheril2015}.  The HDV estimator looks for these perturbations aligned with the background gradient, by taking as input  
two versions of filtered CMB maps.  One map is filtered to keep only large-scales, isolating the smoothly varying background CMB gradient, and we refer to this as the ``large-scale gradient map''. The other map is filtered to keep only small-scales, isolating the perturbations, and we call this the ``small-scale map''.

To minimize foreground bias, we only include CMB multipoles in the range of $\ell \in [5000, 30000]$ in the small-scale map.  Filtering out modes with $\ell < 5000$ in the ``small-scale leg" reduces contributions to the bias from the tSZ signal, which peaks at $\ell \approx 3000$, and the clustering of galaxies, with negligible loss in SNR~\cite{Nguyen2019}.
For the large-scale gradient map, we only include CMB multipoles in the range of $\ell \in [200, 2000]$. In addition, foreground components with known frequency dependence, such as the tSZ, CIB and Radio, can be removed by leveraging multiple frequency channels; the constrained Internal Linear Combination (cILC) technique can deproject these foregrounds from a CMB map with some increase in noise~\cite{Madhavacheril2020, Sailer2021}. However, the deprojection noise penalty is expected to be negligible for the signal dominated large-scale gradient map~\cite{Madhavacheril2018}.  Thus in this work we assume that we can completely remove the tSZ and CIB from the gradient map with negligible increase in noise.\footnote{This deprojection of the large-scale map can be achieved with upcoming Simons Observatory data~\cite{SO2019} with an increase in noise that is below the sample variance uncertainty, and thus negligible.  The removal of the tSZ will be nearly perfect given the known frequency dependence, however, there may be some residual CIB and Radio signal due to uncertainty in the spectral indecies. We discuss this further in Section~\ref{sec:discussion}.}

Since we assume that we can remove non-Gaussian foregrounds in the large-scale gradient map perfectly, that eliminates the primary bispectrum bias and the trispectrum bias terms in Equation~\ref{eq:lens_auto}.  This simplifies Equation~\ref{eq:lens_auto} as follows,\footnote{The factor of four in front of the secondary bispectrum bias in Equation~\ref{eq:lens_auto} disappears in Equation~\ref{eq:lens_auto_cleaned} because the HDV estimator is not symmetric in the two legs.}

\begin{align}
\begin{split}
&\underbrace{\langle \mathcal{Q}[S^{large}_{noFG},S^{small}], \mathcal{Q}[S^{large}_{noFG},S^{small}]  \rangle}_\text{lensing auto spectrum}  \\ &= \underbrace{\langle \mathcal{Q}[T^{large},T^{small}], \mathcal{Q}[T^{large},T^{small}] \rangle}_\text{lensing signal} \\
&+\underbrace{\langle \mathcal{Q}[T^{large},FG^{small}], \mathcal{Q}[T^{large},FG^{small}] \rangle}_\text{secondary bispectrum bias} \\
\end{split}\label{eq:lens_auto_cleaned}
\end{align}
where the only bias term left is the secondary bispectrum bias.

We note that a ``gradient-inversion'' (GI) estimator can also reconstruct the lensing potential on small-scales~\cite{Hadzhiyska2019, Horowitz2019}. In the small-scale, low-noise limit, the GI method should yield even lower lensing reconstruction noise than a QE estimator. We also expect a similar set of foreground mitigation techniques to apply to the GI method; however, since these techniques have been less explored in the context of the GI method, we use the HDV estimator throughout this work.

\subsection{Lensing Reconstruction}

We perform lensing reconstructions on maps that include the lensed CMB temperature and the foreground-reduced maps discussed in Section~\ref{sec:sims}.  The foreground-reduced maps include the non-Gaussian CIB and tSZ components, correlated with the $\kappa$ map and each other, as well as the Radio and reionization kSZ components modelled as Gaussian random fields.  We also add the instrumental white noise levels for CMB-HD, and show the power spectra of these components for 90 and 150 GHz in Figure~\ref{fig:spec} as discussed above.

We coadd the 90~GHz ($M_{90}$) and 150~GHz ($M_{150}$) maps by the inverse of their noise spectra, $N(\ell)_i$ for $i \in [90~{\rm{GHz}},150~{\rm{GHz}}]$, such that

\begin{equation}
M_{coadd}  = \frac{W_{90}}{W_{tot}}M_{90}+ \frac{W_{150}}{W_{tot}}M_{150}
\end{equation}
where $W_i = 1/N(\ell)_i$ and $W_{tot} = \Sigma_i W_i$. We use the publicly available {\it{symlens}} library\footnote{\href{https://github.com/simonsobs/symlens}{https://github.com/simonsobs/symlens}\label{fn:symlens}} to do the lensing reconstructions with and without the non-Gaussian foregrounds added to the maps. We include the Gaussian Radio signal and the reionization kSZ to both the ``with'' and ``without'' non-Gaussian foreground cases; we do not expect either of these to introduce a bias, but we do this to keep consistency between the two cases.  We also include the Radio and reionization kSZ signals in both the large-scale and small-scale maps, anticipating that removing the reionization kSZ will be challenging in general at all scales, and allowing for the increased noise from unresolved Radio sources.

We perform lensing reconstructions in the range of $L \in [5000, 30000]$ using only the lensed CMB temperature maps; at these small scales lensing reconstructions from temperature maps, as opposed to polarization maps, dominate the SNR~\cite{Nguyen2019}. As mentioned above, for the large-scale gradient leg in the HDV estimator, we include CMB multipoles in the range of $\ell \in [200,2000]$.  For the small-scale leg, we use CMB multipoles in the range of $\ell \in [5000,30000]$. We process the simulations described in Section~\ref{sec:sims} to obtain 100 simulated lensing reconstructions.

\subsection{Lensing Auto Spectra}

After generating the lensing reconstructions, we take their power spectra. Since we mask the simulations by a cosine window function ($M^{win}$) before making the reconstructions, we divide the resulting lensing auto spectra by $w_4 = \sum_j{(M^{win}_j)^4} / \sum_j 1$, where $j$ sums over all the pixels in $M^{win}$; this accounts for the mis-normalization of the auto spectra due to the window function. We then subtract the realization dependent lensing bias (RDN0), which arises from the Gaussian component of the auto spectrum~\cite{Namikawa2013}, and the N1 bias, which arises from higher-order corrections to the auto spectrum~\cite{Kesden2003}, from each of the 100 reconstructed auto spectra following~\cite{Story2015, Sherwin2017}. We discuss in detail our RDN0 and N1 computations in Appendix~\ref{sec:rdn0n1}. CMB lensing analyses often further subtract a mean-field $\kappa$ map, which is a transfer function that mainly affects large scales ($L < 100$), from reconstructed $\kappa$ maps; however, since the mean field contribution is negligible compared to the reconstruction noise in the range of $L \in [5000,30000]$, it can be safely ignored. 

While the $\kappa$ realizations (and hence also their power spectra) are not fully independent from each other due to the shared foreground components ($M_{CIB}^{S10}$ and $M_{tSZ}^{S10}$), these simulated spectra are sufficient to compute a covariance matrix. On the scales of $L \in [5000,30000]$, the lensing reconstruction is noise-dominated. Therefore, the sample variance contribution to the total covariance matrix is subdominant to that from the reconstruction noise, $N_L$. Thus, we compute a covariance matrix using the 100 simulated lensing power spectra after subtracting the RDN0 and the N1 biases from each simulation.  

Since the simulations have a sky area of 400 square degrees, we rescale the simulated covariance matrix to match the expected covariance for an observation over half the sky, $f_{sky} = 0.5$.  While the RDN0 subtraction is important in making the covariance matrix roughly diagonal~\cite{Nguyen2019}, the off-diagonal components are not negligible.  We find that the off-diagonal components of our simulation-based covariance matrix are not fully converged after processing 100 simulations. Thus, we replace the off-diagonal elements of our covariance matrix with those computed by~\cite{Nguyen2019}, which work used 1,000 simulations and a lensing estimator and experimental configuration (0.25 arcmin resolution and $0.5~\mu$K-arcmin white noise) nearly identical to that used in our analysis. In addition, that work included the reionization kSZ in the covariance matrix calculation, however, no additional foregrounds. Given that the tSZ and CIB foregrounds contribute only $20\%$ to the diagonals of our covariance matrix, we do not expect them to appreciably contribute to the off-diagonal elements. In Appendix~\ref{sec:kappavariance}, we compare a covariance matrix obtained from an analytic calculation with our simulation-based covariance matrix and discuss the impact of the off-diagonal elements of the covariance matrix on the SNR.

To calculate the bias to the lensing signal from the non-Gaussian correlated tSZ and CIB foregrounds, we compute the lensing auto spectra with and without them, which we label $C_L^{\kappa\kappa, NGFG}$ and $C_L^{\kappa\kappa, GFG}$, respectively. Then, we compute the foreground bias, $C_L^{FG\:\textrm{bias}}$, as the difference between the two lensing auto spectra. For the case ``without'' these correlated non-Gaussian foregrounds (i.e. $C_L^{\kappa\kappa, GFG}$), we match the 2D power spectrum of the non-Gaussian foregrounds patch by patch. To do this, we generate Gaussian tSZ and CIB maps by randomizing the phases of the non-Gaussian tSZ and CIB maps; this phase randomization not only makes the resulting tSZ and CIB maps Gaussian but also breaks the correlations between tSZ, CIB, and $\kappa$. We further improve the match between these two cases by using the same realizations of CMB, $\kappa$, and instrumental noise. These shared realizations, plus the phase-randomized foreground maps, make the RDN0 and N1 biases closely match for the two cases, minimizing any apparent lensing bias due to slight mis-estimations of RDN0 and N1.  Explicitly the lensing auto spectrum bias is given by
 
\begin{figure*}[t]
  \centering
  \includegraphics[width=0.49\textwidth, height=7.7cm]{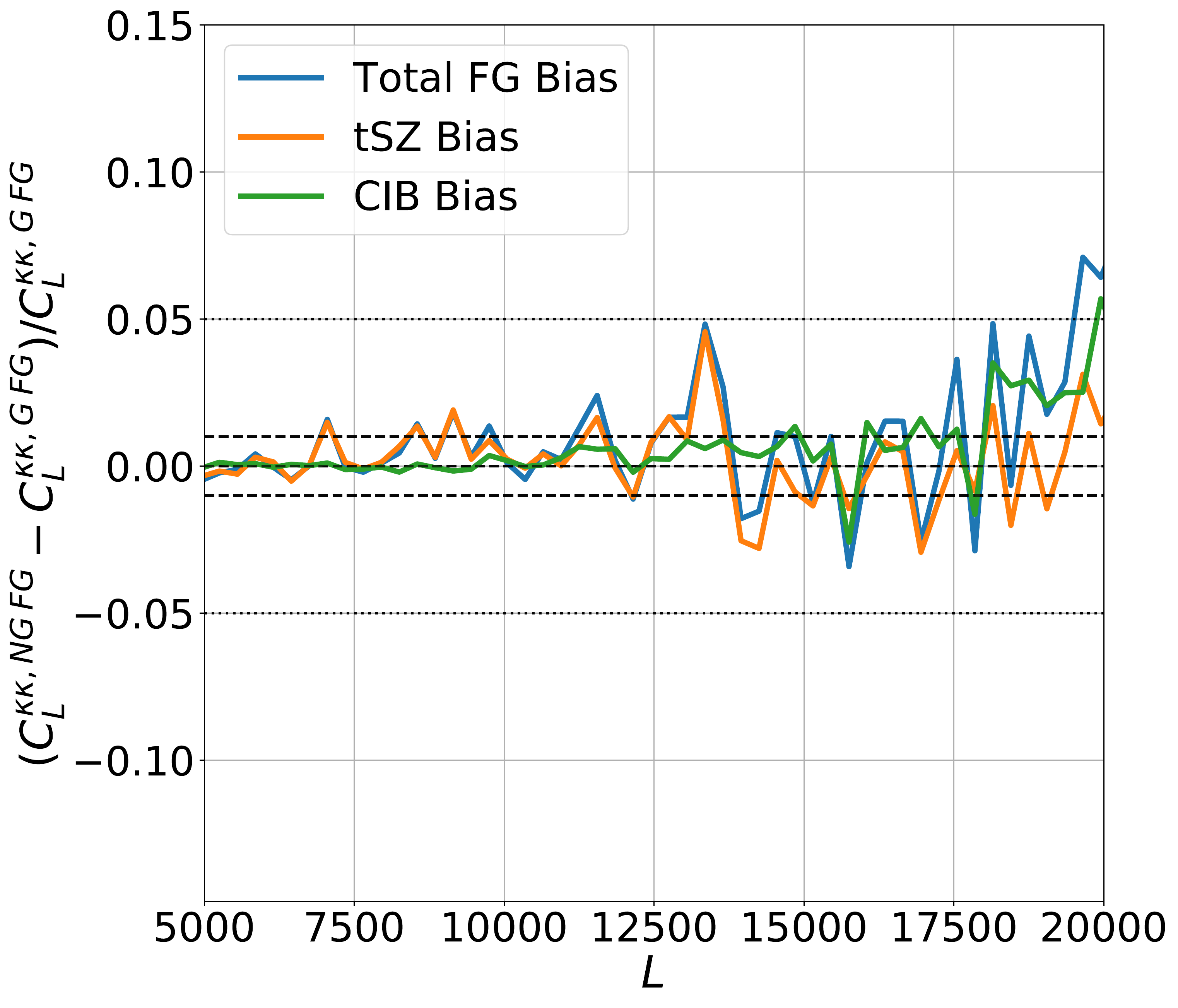}
  \includegraphics[width=0.49\textwidth]{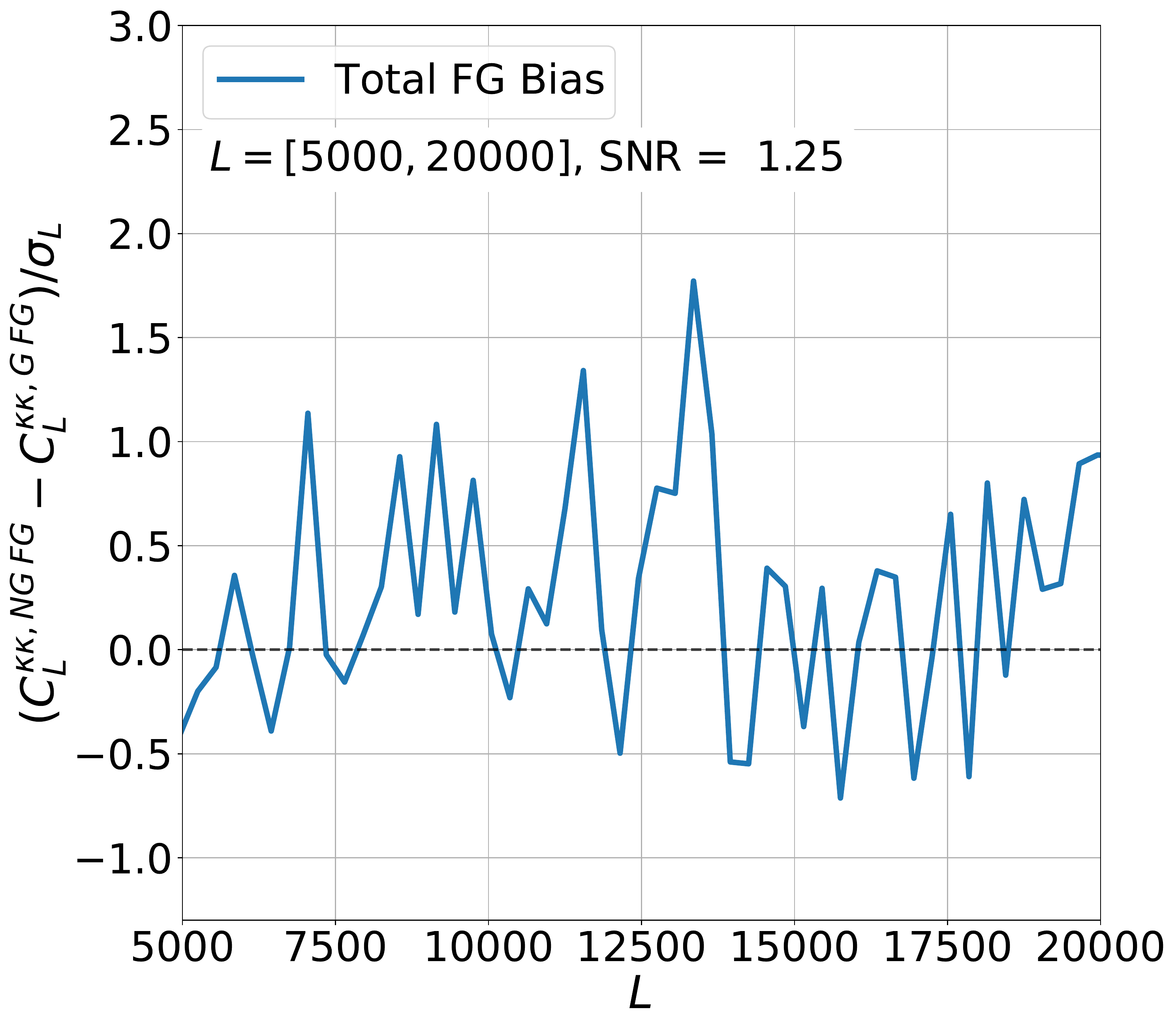}
  \caption{{\it{Left:}} Shown is the fractional lensing bias for various foreground scenarios. The total foreground case (blue curve) includes both tSZ and CIB foregrounds, while the other two cases include either tSZ alone or CIB alone. Here, tSZ and CIB foregrounds are either modelled as uncorrelated Gaussian random fields (G) or prepared from the non-Gaussian (NG) simulations discussed in Section~\ref{sec:sims}.  We include the reionization kSZ and Radio sources, modeled as Gaussian random fields, to all scenarios. The black dashed lines indicate $1\%$ bias, while the black dotted lines indicate $5\%$ bias.  We find that the foreground bias is below a few percent for most of the range $L \in [5000,20000]$.  {\it{Right:}} Shown is the signal-to-noise ratio (SNR) of the total foreground bias (both tSZ and CIB) over half the sky, $f_{sky} = 0.5$. The error bars are the square root of the diagonal elements of the covariance matrix described in Section~\ref{sec:results}. The simulation-based covariance matrix is computed by processing 100 simulations described in Section~\ref{sec:sims} through the lensing reconstruction and auto spectrum pipeline described in Section~\ref{sec:method}. Using the full covariance matrix, we find that the total foreground bias to the lensing auto spectrum deviates from zero by only \fgsnrmissub~in the range of $L \in [4800,20100]$.}
  \label{fig:bias}
\end{figure*} 
 
\begin{equation}
\begin{split}
C_L^{FG\:\textrm{bias}} & =  \frac{1}{100}\sum_{m=1}^{100} ({C_{L,m}^{\kappa\kappa, NGFG} - C_{L,m}^{\kappa\kappa, GFG}}) \\
& =  \frac{1}{100}\sum_{m=1}^{100}[(C_{L,raw,m}^{\kappa\kappa, NGFG}-RDN_{0,m}^{NGFG}-N_1^{NGFG}) \\ 
& - (C_{L,raw,m}^{\kappa\kappa, GFG}-RDN_{0,m}^{GFG}-N_1^{GFG})]. 
\end{split}
\end{equation}
where $m$ sums over the 100 reconstructed auto spectra. Note that the N1 bias term does not have $m$ subscript because the same N1 bias is subtracted from each simulation. Here, any additive bias not induced by the non-Gaussian correlated tSZ and CIB maps will cancel out in the difference. Similarly, any multiplicative bias will cancel out in the ratios used to calculate the fractional bias or SNR, since the denominators are also a function of the simulated auto spectra.

\section{Results}
\label{sec:results}

Figure~\ref{fig:bias} shows the effect of the  extragalactic foregrounds on the lensing reconstruction in the range of $L \in [5000, 20000]$; this $L$ range is chosen because it contributes most of the SNR in determining if there is a problem with CDM on small scales~\cite{Nguyen2019}. Following~\cite{Nguyen2019}, we use a uniform binning of $\Delta \ell =300$.  The left panel shows the fractional lensing bias for various foreground scenarios (CIB alone, tSZ alone, and CIB plus tSZ). Here, the foreground bias is computed as the difference between the reconstructed lensing auto spectrum with and without the non-Gaussian tSZ and/or CIB foregrounds. All cases have the same realizations of CMB, $\kappa$, instrumental noise, and Gaussian Radio and reionization kSZ foregrounds, such that the difference in the lensing spectra is solely due to the tSZ/CIB foregrounds. The total foreground bias (blue curve) includes both CIB and tSZ, and is within a few percent of the unbiased spectra over the range of interest.

The right panel of Figure~\ref{fig:bias} shows the total foreground bias divided by the diagonal error bars of the simulation-based $C_L^{\kappa\kappa}$ covariance matrix discussed in Section~\ref{sec:method}. We find that the total foreground bias is below $0.5\sigma$ for many spectral bins in the range shown. In order to determine whether this foreground bias is detectable, we use the full covariance matrix to calculate the SNR as follows. The $\chi^2$ of the total foreground bias with respective to a null signal (i.e., $C_L^{bias}$=0) is computed as 
\begin{equation}
{\chi^2_{\textrm{null}}} = \sqrt{\sum_{L,{L'}} (X_L - Y_L)C^{-1}_{L,{L'}}(X_{L'} - Y_{L'})} \label{eq:snr_fg}
\end{equation}
where $X_L = C_{L}^{\kappa\kappa, NG FG}-C_{L}^{\kappa\kappa, G FG}$, $Y_L =0$, and $C^{-1}_{L,{L'}}$ are the elements of the inverted covariance matrix. We calculate this over the range of $L \in [4800,20100]$.  
We compute $\chi^2_{\textrm{best-fit}}$ by replacing $Y_L=0$ with the best-fit $C_L^{bias}$ line obtained from fitting a straight line in the range of $L \in [4800,20100]$; for the total FG case we find a slope of $7.91 \times 10^{-17}$ and an intercept of $2.74 \times 10^{-14}$ for this best-fit. We also obtain best-fit lines for the tSZ-only and CIB-only cases. We find that the lensing bias in the total FG case favors the best-fit line over the null at only \fgsnrmissub, where we compute this significance as $\sqrt{\chi_{\textrm{null}}^2-\chi_{\textrm{best-fit}}^2}$.  We find that the significance of detecting the foreground bias for the tSZ-only and CIB-only cases is \fgsnrtszmissub~and \fgsnrcibmissub, respectively. As discussed in more detail in Appendix~\ref{sec:kappavariance}, the inclusion of the off-diagonal elements of the covariance matrix is important for the total SNR calculation. We find that the total SNR would increase by a factor of two if we used only the diagonal elements to compute the SNR.

\begin{figure}[t]
  \centering
  \hspace{-3mm}\includegraphics[width=0.97\columnwidth]{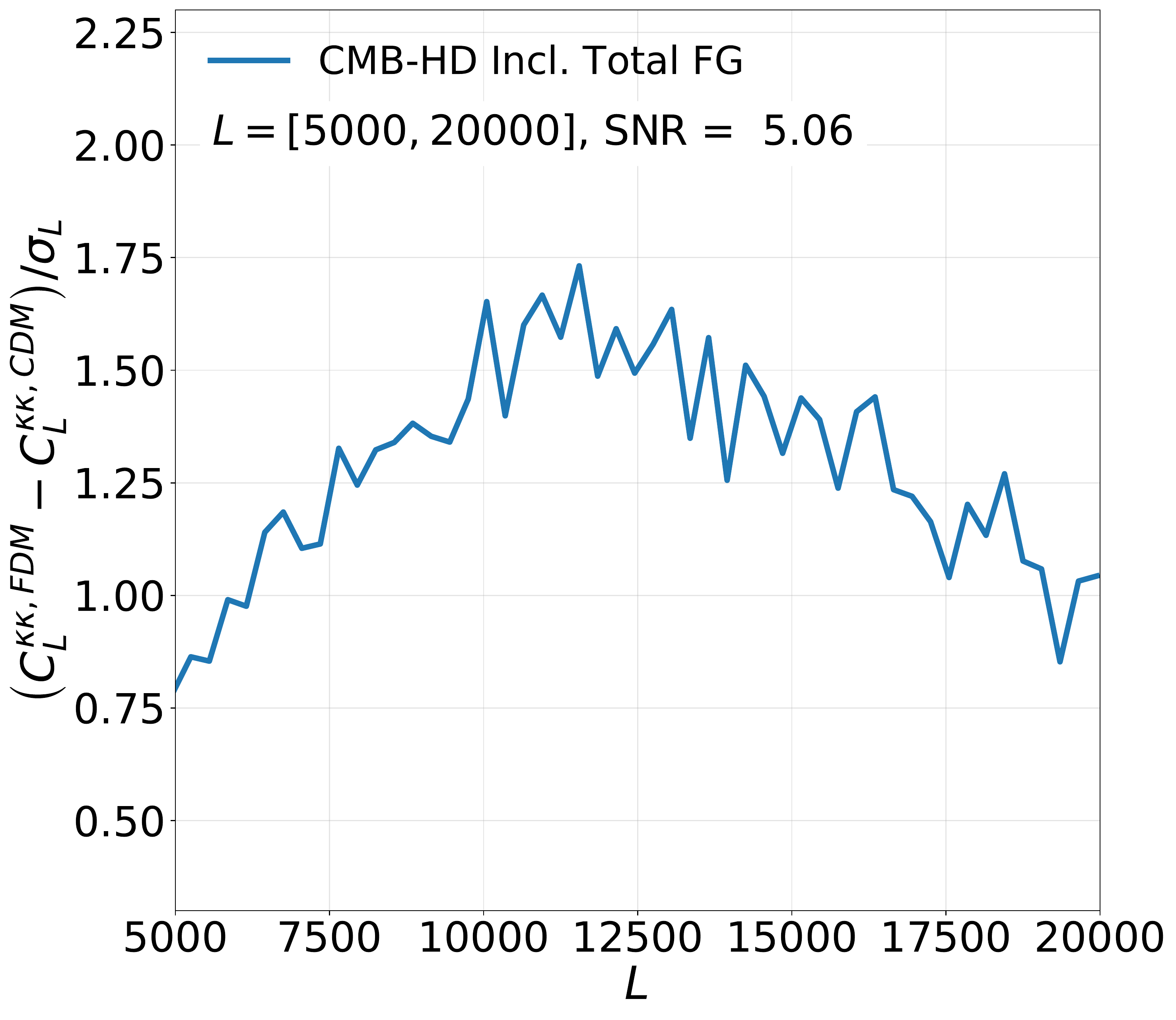}
  \caption{Shown is the SNR for distinguishing an $m = 10^{-22}$~eV FDM model from a CDM model when including the noise from extragalactic foregrounds in the covariance matrix. The foregrounds include the reionization kSZ, tSZ, CIB, and Radio sources, with the source cuts discussed in Section~\ref{sec:sims}, resulting in the power spectra shown in Figure~\ref{fig:spec}. Here, the signal is the difference between a theoretical  $10^{-22}$~eV FDM $C_\ell^{\kappa \kappa}$ and a theoretical CDM $C_\ell^{\kappa \kappa}$ obtained from the WarmAndFuzzy code~\cite{Marsh2016a}. We do not include bias from the foregrounds in either the CDM or FDM models here. The error bars are the square root of the diagonals of the covariance matrix described in Section~\ref{sec:method}.  Using the full covariance matrix, we find a SNR of \fdmsnrmissub~in the range of $L \in [4800,20100]$. As shown in the right panel of  Figure~\ref{fig:bias}, the foreground bias is not detectable in this same $L$ range.}
  \label{fig:snr}
\end{figure}

Figure~\ref{fig:snr} shows the SNR for distinguishing between an $m = 10^{-22}$~eV FDM model~\cite{Hui2017, Hu2000a} and a CDM model, in the presence of extragalactic foregrounds. Here, the signal is the difference between a theoretical $m = 10^{-22}$~eV FDM $C_\ell^{\kappa \kappa}$ and a theoretical CDM $C_\ell^{\kappa \kappa}$ obtained from the WarmAndFuzzy code~\cite{Marsh2016a} and used in~\cite{Nguyen2019}. Here we use the same error bars and covariance matrix used in the right panel of Figure~\ref{fig:bias}.  We compute the total SNR in the range of $L \in [4800,20100]$ following 

\begin{equation}
{\rm{SNR}} = \sqrt{\sum_{L,{L'}} (X_L - Y_L)C^{-1}_{L,{L'}}(X_{L'} - Y_{L'})} \label{eq:snr_fdm}
\end{equation}
where $X_L = C_{L}^{\kappa\kappa, FDM}$, $Y_L = C_{L}^{\kappa\kappa, CDM}$ (the FDM and CDM $C_{L}^{\kappa\kappa}$, respectively), and $C^{-1}_{L,{L'}}$ are the elements of the inverted covariance matrix, including off-diagonal elements.  Calculating this, we obtain a SNR of \fdmsnrmissub. As shown in the right panel of Figure~\ref{fig:bias}, the foreground bias in the same $L$ range is not detectable.

\begin{figure}[t]
  \centering
  \includegraphics[width=0.49\textwidth]{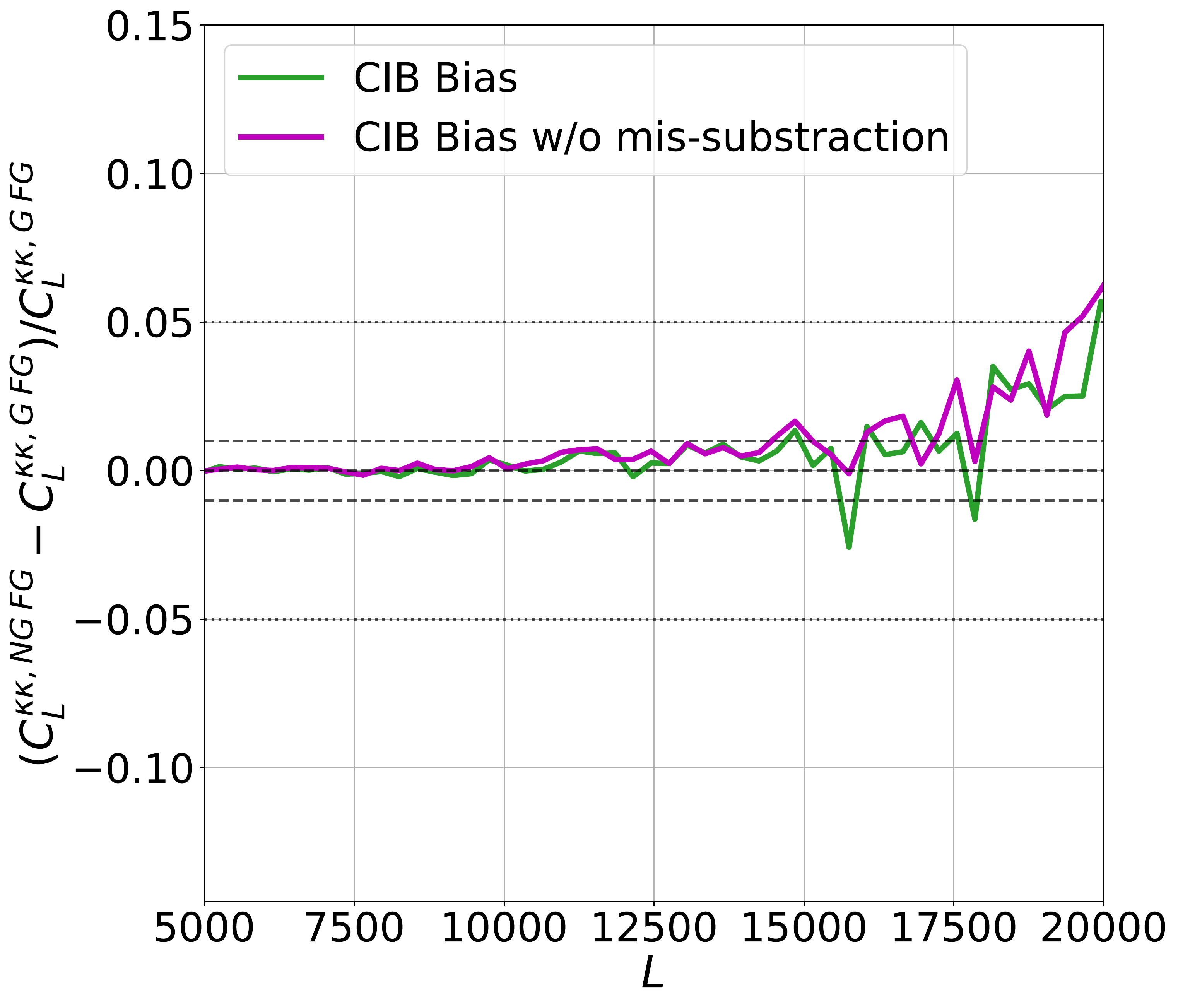}
  \caption{Shown is the fractional lensing bias due to the CIB foreground with and without accounting for some uncertainty in the CIB spectral indecies. The black dashed lines indicate 1\% bias, while the black dotted lines show 5\% bias. The magenta curve represents the case where we can perfectly remove all the CIB sources in the range of 0.03 mJy to 0.2 mJy at 150 GHz. The green curve shows the case where some sources are either over or under subtracted by up to 5\% of their original fluxes in both 90 and 150 GHz maps.  After including the effect of the mis-subtraction, the SNR of the total foreground bias (including both CIB and tSZ) in the range of $L \in [4800, 20100]$ gets smaller,  changing from \fgsnr~to \fgsnrmissub~(see Section~\ref{sec:results} for details). The total SNR in distinguishing a $10^{-22}$ eV FDM model from a CDM model changes only slightly from \fdmsnr~to \fdmsnrmissub.}
  \label{fig:bias_cib_missub}
  \vspace{2mm}
\end{figure}

\subsection{Uncertainty in the CIB Spectral Index}

Even though CMB-HD will detect CIB sources with an unprecedented sensitivity~\cite{Sehgal2019a, Sehgal2019b, Sehgal2020}, there will likely be some uncertainty in the CIB spectral indecies for the fainter sources. Since we template-subtract some dimmer CIB sources by extrapolating their flux measurements at the 280~GHz channel down to 150 and 90 GHz, this uncertainty can lead to a mis-subtraction of the CIB sources. In our baseline analysis, we include this mis-substraction effect (see Section~\ref{sec:sims} for details). Here, we quantify its effect on the foreground bias and the simulation-based covariance matrix.

In our baseline analysis, we ``mis-subtract'' CIB sources in the flux range of 0.03~mJy to 0.2~mJy at 150 GHz, in both 90 GHz and 150 GHz maps by up to 5\%.  These sources will be detected in 280 GHz maps with a significance between 5 and about $30\sigma$, and thus will have some uncertainty from the flux measurement itself at 280 GHz as well as some uncertainty in the CIB spectral index when extrapolating to 150 GHz. We pick a mis-subtraction level up to $5\%$ for all the sources, noting that for the bright sources, $5\%$ mis-subtraction may be conservative since their fluxes are measured with better precision than that, and they will also be detected at 220 GHz, thereby lowering the uncertainty in their spectral index; for fainter sources, a larger mis-subtraction amount may be warranted, however, since these sources are faint, that should minimally increase the noise, and as we show below, may serve to lower the lensing auto spectrum bias.       

We also run a case where these CIB sources are subtracted perfectly and their pixel values replaced with the mean flux of the residual CIB sources. Figure~\ref{fig:bias_cib_missub} shows the effect of this CIB mis-subtraction compared to the perfect subtraction case.  Overall, the lensing bias from the CIB component stays roughly the same in the range of $L \in [4800,20100]$. When including the effect of mis-subtracting the CIB sources, the SNR of the total FG bias goes from \fgsnr~to \fgsnrmissub. The reduction in the SNR is likely due to both the increase in the lensing reconstruction noise and the decrease in the CIB foreground bias. For the latter, the random ``mis-subtraction'' may break some correlation between the $\kappa$ and CIB maps resulting in the lower foreground bias. The effect of the mis-subtraction on distinguishing a $10^{-22}$~eV FDM model from a CDM model (see Figure~\ref{fig:snr}) is also negligible. In this case, the total SNR goes from \fdmsnr~to \fdmsnrmissub~when we include CIB mis-subtraction in our simulations.

\section{Discussion}
\label{sec:discussion}

The results above suggest that the two most dominant foregrounds, the tSZ and the CIB, can be reduced to a level that will not significantly bias the lensing auto spectrum in the range $L \in [4800, 20100]$. In addition, the noise penalty in using the HDV estimator and including the residual foregrounds still allows for the ability to determine if CDM is a good match to the small-scale matter power spectrum, or if structure is suppressed at the level, for example, that a $10^{-22}$~eV FDM model would predict.  

Below we outline some additional foreground contributions that were not included in this analysis, and we outline how we foresee them being mitigated.  In future work, these additional contributions will be explored in more detail.

\subsection{Foregrounds in the Large-scale Gradient Map}

In the current analysis, we assumed that there were no tSZ and CIB foregrounds in the large-scale gradient map.  For this map, we only included CMB multipoles in the range of $\ell \in [200, 2000]$.  We anticipate that these large-scale gradient maps will be provided by SO data, which extends down to CMB multipoles of $\ell=30$~\cite{SO2019}, as opposed to $\ell_{\rm{min}}=1000$ for CMB-HD~\cite{Sehgal2019a}.  Since the frequency dependence of the tSZ signal is exactly known, we can deproject it using the constrained Internal Linear Combination (cILC) technique~\cite{Madhavacheril2020, Sailer2021}.    While there is some noise penalty in using this deprojection technique, given the SO noise levels, the temperature maps will still be sample variance limited over half the sky for $\ell \in (200, 2000)$; the increase in noise from the deprojection will be negligible compared to the uncertainty from the sample variance~\cite{Madhavacheril2018}.

CIB and Radio sources can be similarly deprojected from the large-scale gradient map.  However, uncertainty in the spectral indecies will leave some residual foregrounds.  A path forward would be to template-subtract the same sources detected in CMB-HD maps from the SO maps, and then deproject the SO maps using their multi-frequency data.  In addition, CCAT-p data, which is designed to map the CIB on large-scales, can be used to further clean the CIB from the gradient leg~\cite{CCAT2019}.  Given that our current analysis shows that some amount of source mis-subtraction actually lowers the bias to the lensing auto spectrum, we do not expect residual CIB and Radio sources in the gradient map to add significantly to the lensing bias.  In addition, we can apply point-source hardening techniques to our lensing reconstruction procedure; point-source hardening is shown to reduce the foreground bias from point sources with only small increases in the reconstruction noise~\cite{Sailer2020, Sailer2021}.  We leave a detailed analysis of this to future work.

\subsection{Bias from Radio Galaxies}

We did not explicitly include correlated radio sources in this analysis since we focused on the bias from the tSZ and CIB foregrounds, anticipating these to be the dominant sources of bias as is the case for the lensing auto spectrum on large scales~(see e.g.,~\cite{Han2021}).  As mentioned in Section~\ref{sec:sims}, we expect to be able to template-subtract Radio sources above 0.04 mJy in 90 GHz maps at the $5\sigma$ level.  (Radio sources are subdominant in 150 GHz maps compared to the CIB (see Figure~\ref{fig:spec}).)  Given this low flux cut, and the fact that Radio sources are fewer in number and less correlated with the $\kappa$ map, we do not anticipate significant bias to the lensing auto spectrum from Radio sources; however, we leave a detailed investigation to future work.

\subsection{Bias from the Late-time kSZ Effect}

The kSZ effect arises when CMB photons inverse-Compton scatter off ionized gas that has a bulk motion. As discussed in Section~\ref{sec:sims}, the kSZ effect has two components, 1)~one from the epoch of reionization, and 2)~one from the lower redshift Universe~(e.g.,~\cite{Smith2018}).  We assume in this work that the reionization kSZ, which is from $z \approx 10$, is Gaussian and uncorrelated with the $\kappa$ map, whose kernel peaks at much lower redshift ($z \approx 2$).  We include the reionization kSZ in both the large-scale gradient map and the small-scale map, to add it as an extra noise source; from Figure~\ref{fig:spec}, we see that this is a significant source of noise below $\ell=10000$.  In terms of bias to the lensing auto spectrum, the reionization kSZ has been shown to lead to negligible bias on larger scales than studied here~\cite{Cai2021}, and future investigation is warranted to explore its effect on smaller scales.  

The late-time kSZ may lead to considerable bias to the lensing auto spectrum.  However, there are a number of promising avenues for removing it.  For the large-scale gradient map, removing the late-time kSZ with an overlapping galaxy survey (e.g., the Vera C. Rubin Observatory~\cite{LSST2019}) using variations of the techniques discussed in \cite{Schaan2016, Smith2018, Schaan2021, Cayuso2021}, seems promising. For the small-scale map, recent advances in Deep Learning pattern recognition is a promising avenue to remove the late-time kSZ. While the kSZ effect has no frequency dependence, making it hard to remove with multi-frequency techniques~\cite{Ferraro2018}, the kSZ signal has a characteristic spatial pattern that can be recognized by a convolution neural network (e.g.~\cite{Wu2019, Han2021,  Guzman2021a,  Guzman2021b, Lin2021}). In turn, we can use this recognized spatial pattern to remove the kSZ signal, especially on small scales above $\ell=5000$ where it is the dominant signal. Examples of deep learning foreground mitigation methods are shown in~\cite{Petroff2020,Villanueva2021}. 

Since the kSZ signal is proportional to the bulk motion of the electron gas, and because the motion of the gas toward and away from an observer is equally likely, the mean correlation between the kSZ signal and $\kappa$ is zero. As a result, we expect the secondary bispectrum bias (see Equation~\ref{eq:lens_auto}), which arises due to the correlation between a non-Gaussian foreground and $\kappa$, to be small.  If we can clean the late-time kSZ from either the large-scale or small-scale maps, then this secondary bispectrum bias is the only bias term that will remain.  We leave a thorough examination of the bias from the late-time kSZ to subsequent work.

\section{Conclusion}
\label{sec:conclusion}

In this work, we provide a proof of concept demonstration that lensing bias from the two most dominant foregrounds, the tSZ and CIB, can be mitigated, and that CMB-HD can measure the CMB lensing auto spectrum on small scales ($L \in [5000,20000]$).  We find that the combined tSZ and CIB foreground bias, in the coadded CMB-HD 90 and 150~GHz channels, has a SNR of only $1.3\sigma$.  We also find that modest mis-subtraction of faint CIB sources due to their uncertain spectral indecies has negligible impact.  Using an interesting alternative dark matter model as a benchmark, we find that CMB-HD can distinguish a $10^{-22}$~eV FDM model from a CDM model at the $5\sigma$~level. While this analysis provides an initial proof of concept demonstration, a full investigation including all potential sources of bias, and an exploration of the optimal path to mitigate them is warranted.  
Such measurements of the high-resolution CMB lensing auto spectrum will probe the matter spectrum power spectrum out to small scales ($k \sim 10~h$Mpc$^{-1}$).  In addition to providing a robust measurement of the small-scale matter distribution via gravitational lensing, this will put stringent constraint on alternative dark matter models and baryonic processes, informing both fundamental physics and galaxy evolution.


\begin{acknowledgments}
DH and NS acknowledge support from DOE grant number DE-SC0020441. DH and NS also thank Gil Holder, Mathew Madhavacheril, Blake Sherwin, and Alexander van Engelen for useful comments.  NS warmly thanks the Aspen Center for Physics, which is supported by National Science Foundation grant PHY-1607611, for their hospitality in September 2021, where early ideas for this work were informally discussed.
\end{acknowledgments}

\appendix

\section{Gaussian and non-Gaussian $\kappa$ maps}\label{sec:gvsngkappa}

\begin{figure}[t]
  \centering
  \includegraphics[width=0.45\textwidth]{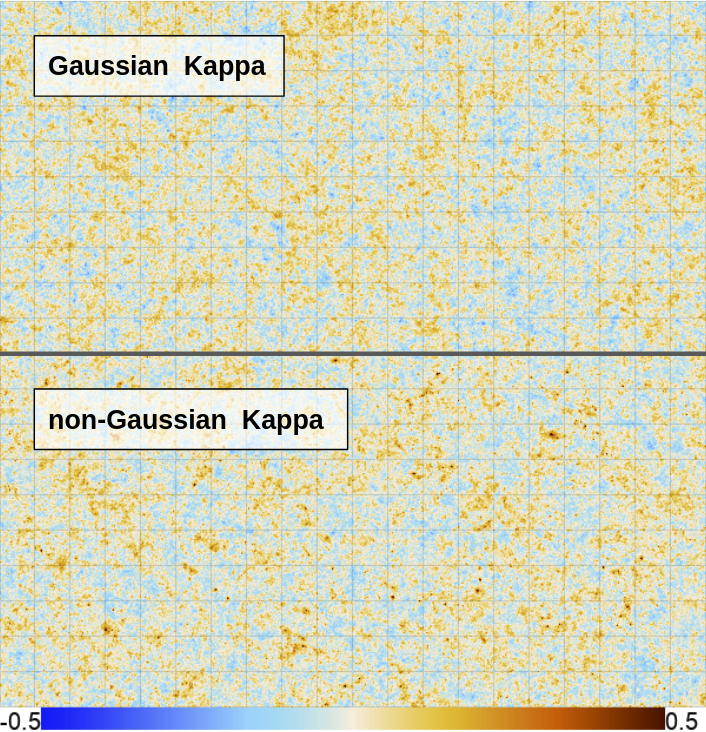}
  \caption{Shown is a Gaussian (top panel) and a non-Gaussian (bottom panel) $\kappa$ map. We generate the non-Gaussian $\kappa$ map using Equation~\ref{eq:kappasim}, and the Gaussian $\kappa$ map as a Gaussian random field with the same $C_{\ell}^{\kappa \kappa}$. By construction, the non-Gaussian $\kappa$ map is 70\% correlated with the S10 CIB map and 50\% correlated with the S10 tSZ map at 150~GHz.  Visually, we find that the large-scale fluctuations look similar between the Gaussian and non-Gaussian $\kappa$ maps, however, there is more small-scale structure in the non-Gaussian $\kappa$ map that mirrors bright structures in the CIB and tSZ maps.  We expect that these $\kappa$ maps will yield a bias to the lensing auto spectrum that is conservative, as discussed in Section~\ref{sec:sims}.}\label{fig:kappa}
\end{figure}

In Figure~\ref{fig:kappa}, we show a visual comparison between one realization of the non-Gaussian $\kappa$ map generated via the procedure discussed in Section~\ref{sec:sims}, and a Gaussian $\kappa$ map generated with the same $C_{\ell}^{\kappa \kappa}$. We generate the Gaussian $\kappa$ map as a Gaussian random field, and use many realizations of them to compute a realization dependent bias (RDN0) and N1 bias (see Section~\ref{sec:method} and Appendix~\ref{sec:rdn0n1} for details). The non-Gaussian $\kappa$ map is generated using Equation~\ref{eq:kappasim} from Section~\ref{sec:sims}, and is 70\% correlated with the S10 CIB map and 50\% correlated with the S10 tSZ map at 150~GHz by construction.  Note that Equation~\ref{eq:kappasim} leaves in the maximum amount of non-Gaussian structure from the S10 CIB and tSZ maps in the non-Gaussian $\kappa$ maps. In general, this will lead to larger higher-order correlations between these non-Gaussian foregrounds and the non-Gaussian $\kappa$ map than expected. Visually, we find that the large-scale fluctuations look similar between the Gaussian and non-Gaussian $\kappa$ maps. However, we find more small structure in the non-Gaussian $\kappa$ maps. In particular, the brighter spots in the non-Gaussian $\kappa$ maps roughly coincide with the location of bright CIB sources.

\section{Simulation-based RDN0 and N1 Computations}\label{sec:rdn0n1}
We compute the realization dependent Gaussian lensing bias (RDN0)~\cite{Namikawa2013} following Equation 7 from~\cite{Sherwin2017}. Unlike the analytic N0 bias described in~\cite{Hu2002}, the RDN0 bias was designed to account for a slight difference between the power spectra of simulations and data. In this work, our ``data'' is the realistic non-Gaussian and phase-randomized simulations discussed in Section~\ref{sec:sims}. Our ``simulations'' are Gaussian simulations constructed to match the power spectra of these realistic simulations. 

\begin{figure}[t]
  \centering
  \includegraphics[width=0.49\textwidth]{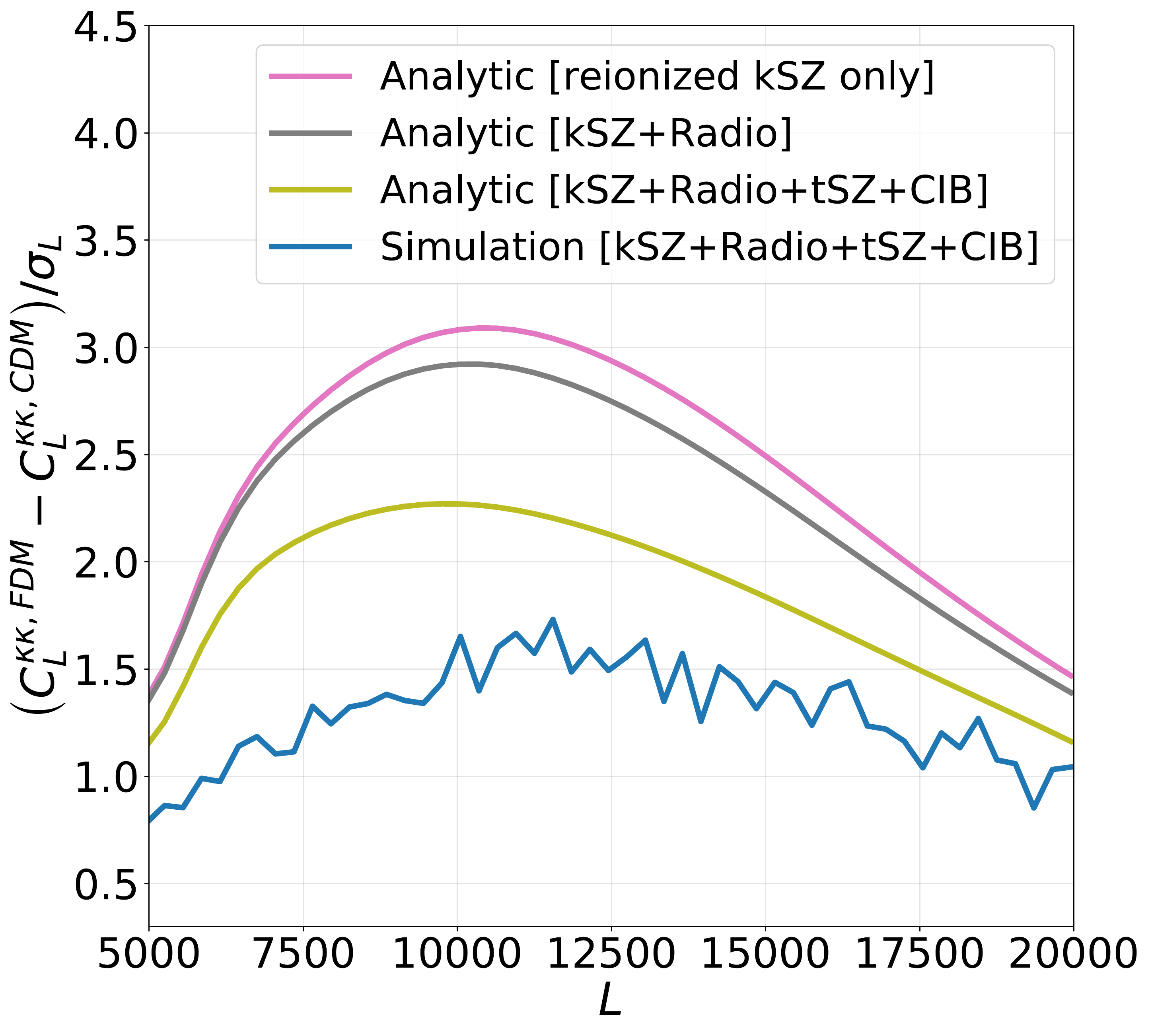}
  \caption{The same as in Figure~\ref{fig:snr}, except for three additional curves calculated with analytic error bars. The blue curve is the same as shown in Figure~\ref{fig:snr}. We compute the analytic error bars for other three curves using Equation~4 from~\cite{Knox1995} given the theoretical $C_L^{\kappa\kappa}$ spectrum and analytic N0 lensing noise curves.  The same experimental configuration (white noise level, resolution, and sky coverage) is assumed in all four cases. As in~\cite{Nguyen2019}, we find that the diagonal variance from simulations is larger than that predicted by the analytic calculation. We also find that the off-diagonal variance has a large impact on the total SNR calculation. Using only the diagonal variance, we find that the analytic case (yellow curve) gives a SNR of \fdmsnranalyticdiag, while the simulation-based case (blue curve) gives a SNR of \fdmsnrmissubdiagonly. Once we include the off-diagonal variance for the simulation-based case, we find that the SNR reduces by roughly a factor two, going from \fdmsnrmissubdiagonly~to \fdmsnrmissub.}\label{fig:SNR-comparison}
\end{figure}

We calculate the RDN0 bias as
\begin{align}\label{eq:rdn0}
\begin{split}
& C_L^{RDN0}[XY,AB] \\
& =  \langle \mathcal{Q}[XY^S,AB^S] + \mathcal{Q}[XY^S,A^SB] \\
& + \mathcal{Q}[X^SY, AB^S] + \mathcal{Q}[X^SY, A^SB] \\
& - \mathcal{Q}[X^SY^{S'}, A^{S}B^{S'}] - \mathcal{Q}[X^SY^{S'}, A^{S'}B^{S}]\rangle_{S,S'}
\end{split}
\end{align}
where $\mathcal{Q}$ stands for a quadratic estimator.  Since Equation~\ref{eq:rdn0} does not assume any symmetry in $\mathcal{Q}$, it works for the HDV estimator without any modification. $S$ and $S'$ superscripts indicate two independent realizations of Gaussian simulations. For each of the 100 ``data'' simulations, we average over 20 RDN0 computations, where each RDN0 run requires 2 new Gaussian simulations (for $S$ and $S'$); thus, in total, we make $2 \times 100 \times 20 = 4000$ Gaussian simulations for the RDN0 calculation.   
As in~\cite{Nguyen2019}, we find that subtracting the RDN0 bias removes a large amount off-diagonal correlation in the simulation-based $C_L^{\kappa \kappa}$ covariance matrix. 

Similarly, we compute the N1 bias, which arises from higher-order corrections to the auto spectrum~\cite{Kesden2003}, following~\cite{Story2015, Sherwin2017}. 

\begin{align}\label{eq:n1}
\begin{split}
& C_L^{N1}[XY,AB] \\
& =  \langle \mathcal{Q}[X^{S_{\phi}}Y^{S_{\phi}^{'}},A^{S_{\phi}}B^{S_{\phi}^{'}}] + \mathcal{Q}[X^{S_{\phi}}Y^{S_{\phi}^{'}},A^{S_{\phi}^{'}}B^{S_{\phi}}] \\
& - \mathcal{Q}[X^SY^{S'}, A^{S}B^{S'}] - \mathcal{Q}[X^SY^{S'}, A^{S'}B^{S}]\rangle_{S,S',S_{\phi},S_{\phi}^{'}}
\end{split}
\end{align}
Here, $S$ and $S'$ superscripts again indicate two independent realizations of Gaussian simulations, while $S_{\phi}$ and $S_{\phi}^{'}$ superscript stand for two independent CMB realizations lensed by the same $\kappa$ map. Note that the N1 bias is computed using only the Gaussian simulations. We average over 100 N1 computations; using, in total, $4\times 100 = 400$ Gaussian simulations.

\section{Analytic versus Simulation-based $C_L^{KK}$ Variance}
\label{sec:kappavariance}

Figure~\ref{fig:SNR-comparison} shows the comparison of SNRs for distinguishing a $10^{-22}$~eV FDM model from a CDM model, computed either using an analytic or simulation-based $C_L^{KK}$ covariance matrix. The error bars for the blue curve are the square root of the diagonal elements of the simulation-based covariance matrix described in Section~\ref{sec:results}. We compute the analytic error bars for the other three curves using Equation~4 from~\cite{Knox1995} given the theoretical $C_L^{\kappa\kappa}$ spectrum and analytic N0 lensing noise curves. The analytic N0 lensing noise curves are computed using {\it{symlens}} assuming the same experimental configuration (white noise level, resolution, and sky coverage) used in the simulated case. As in~\cite{Nguyen2019}, we find that the diagonal variance from the simulations is larger than that predicted by the analytic calculation (compare yellow and blue curves). Where the discrepancy between the diagonal variance is largest coincides with where the simulation-based $C_L^{\kappa \kappa}$ band powers are significantly correlated $L \in [5000,12000]$. In addition, we find that the off-diagonal variance has a large impact on the total SNR. Using only the diagonal of variance, we find that the analytic case (yellow curve) gives a SNR of \fdmsnranalyticdiag, whereas the simulation-based case (blue curve) gives a SNR of \fdmsnrmissubdiagonly. However, once we include the off-diagonal variance for the simulation-based case, we find that the SNR reduces by roughly a factor of two, going from \fdmsnrmissubdiagonly~to \fdmsnrmissub. Note that the analytic variance calculations do not include any off-diagonal contributions.  We also show how the analytic diagonal variance changes with the addition of different foregrounds.

\bibliography{main.bib}

\end{document}